# Coulomb blockade and quantum confinement in field electron emission from heterostructured nanotips

Victor I. Kleshch[1,*], Vitali Porshyn[2], Dirk Lützenkirchen-Hecht[2], Alexander N. Obraztsov[1,3]

[1]*Department of Physics, Lomonosov Moscow State University, Moscow 119991, Russia*

[2]*Physics Department, Faculty of Mathematics and Natural Sciences, University of Wuppertal, Wuppertal 42119, Germany*

[3]*Department of Physics and Mathematics, University of Eastern Finland, Joensuu 80101, Finland*

* Corresponding author, e-mail: klesch@polly.phys.msu.ru

A new field electron emission (FE) mechanism, which includes Coulomb blockade and quantum confinement effects, is revealed for heterostructured emitters composed of quantum dots and nanowires self-assembled at diamond nanotips. The total energy distributions of the emitted electrons show multiple peaks attributed to the discrete electronic states of the quantum-confined emitter with the corresponding energy levels oscillating as a function of the applied voltage due to the Coulomb blockade. The FE current-voltage characteristics exhibit a modified Coulomb staircase with additional steps becoming more pronounced with increasing voltage. The experimentally observed behavior is consistent with numerical simulations based on the model of Coulomb blockade in quantum dots in combination with the theory of FE from sharp tips.

The Coulomb blockade (CB) and quantum confinement effects in nanoscale heterostructures make it possible to manipulate individual electrons providing a platform for fundamental studies in quantum coherent electronics [1] and metrology [2]. In solid-state heterostructures, e.g. in a two-terminal system consisting of a nanostructure electrically isolated from the source and drain leads by tunnel junctions, potential barrier profiles are usually insensitive to the number, $N$, of electrons determining the charge, *Ne*, of the nanostructure. However, this is not the case in a vacuum electronic device based on field emission (FE) where the barrier between the nanostructure and the drain (anode) has a triangular shape which depends on the field strength proportional to *N*. This alters the physics of charge transport and allows observations of single-electron charging effects in FE, such as a Coulomb staircase in the current-voltage characteristics, under conditions which cannot be realized in solid-state devices [3].

Several experimental observations of FE from individual nanostructures [4-6] and molecules [7] in the single-electron regime have been reported over the last decade. The most remarkable behavior (in comparison to solid-state devices) was observed for nanostructures attached to a needle-shaped cathode and located at a macroscopic distance from the anode. In particular, the Coulomb staircase was observed for temperatures of up to 1000 K, operating currents up to several microamperes, and voltages, necessary to transfer an extra electron to the nanostructure, of few hundreds of volts [4,6]. In these pioneering experiments single-electron charging was observed for carbon nanotubes and nanowires where quantum confinement effects are small and are not manifested in FE. Here we make the next step by exploring FE in heterostructures based on small-size quantum-confined carbon nanowires. The heterostructures were created by modification of the atomic structure of the diamond needle-like crystallites caused by the action of a strong electric field and Joule heating during FE [6]. First, we describe the main stages of this FE-induced reconstruction process (Fig. 1) and then present the corresponding experimental and simulation results.

Initially, at low FE currents, nanoscale structures are formed at the apex of a pristine (unmodified) diamond nanotip due to the electric field-assisted diffusion of surface atoms. Self-organization of such kind of nanostructures is a well-known phenomenon in FE from uncleaned emitters, e.g. carbon nanotubes [8,9], graphene [10], tungsten tips [11,12] and Spindt cathodes [13]. In this case, electrons tunnel into the vacuum through discrete energy states (separated by a characteristic energy, $\Delta\varepsilon$, see Fig. 1a) arising due to the quantum confinement in the nanostructure, which can therefore be referred to as a quantum dot (QD). At high FE currents, QDs evaporate, and the surface diamond layer with a few-nanometer thickness transforms into amorphous carbon (a-C). After that, electron emission is typically governed by the classical Fowler-Nordheim (FN) mechanism [14,15] (Fig. 1b). Furthermore, elongated carbon nanostructures (nanowires) with a length of about 10 nm can be formed by varying the electric field and FE current as directly observed by transmission electron microscopy [6]. The quantum confinement is small and has no effect on FE in such nanowires. However, since the nanowire is electrically isolated from the underlying low-conductive a-C layer by a potential barrier (formed due to the difference in sp2-hybridized carbon content), CB of the FE current (characterized by the charging energy, $\delta\varepsilon$, see Fig. 1c) becomes observable. Finally, one can imagine a structure where CB and quantum confinement coexist. In the case of QD structure formation directly at the apex of a nanowire, a double-well structure is formed (Fig. 1d), and single-electron charging and energy-quantization effects act independently on charge transport. Also, the electron quantum confinement is possible in the nanowire itself (Fig. 1e), when its length is sufficiently small, i.e. when it is comparable to the de Broglie wavelength. In this case, an interplay between CB and quantum confinement may be observed since both effects coexist in one and the same structure.

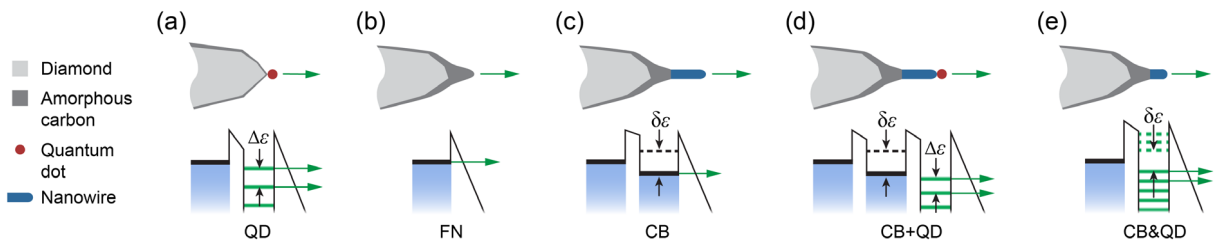

**FIG. 1.** (a-e) Schematic illustration of the emitter and corresponding energy diagrams at different stages of the FE-induced structure modification. See text for details.

The FE experiments were carried out at room temperature in an ultrahigh vacuum setup using a diode configuration with a planar metal mesh anode and a cathode consisting of a diamond needle attached to a tungsten support tip. The DC voltage, $V$, was applied between the electrodes. The total FE current, $I$, was measured by a picoammeter, and the spectrum (total electron energy distribution), $J(\varepsilon)$, was measured by an electron spectrometer with a hemispherical analyzer located behind the mesh anode. The experimental details can be found elsewhere [6,16].

A set of representative measurements for one of the diamond needles at different stages of the FE-induced reconstruction process is shown in Figure 2. The initially unstable FE was stabilized after reaching the current of about 100 nA (see Supplemental Material [17]). An example of well reproducible measurements obtained after stabilization is shown in Fig. 2a, where we present the dependence of $J(\varepsilon)$ on $V$, the corresponding $I(V)$ curve and the position of the energy peaks, $\varepsilon_{peak}(V)$, for each $J(\varepsilon)$ curve. The spectrum contains two well-separated peaks, which linearly shift with $V$. Moreover, a third peak appears at high $V$. This behavior is in good agreement with FE model involving localized states of a self-assembled QD structure (Fig. 1a). The energy peaks correspond to discrete energy states shifting with voltage due to penetration of the electric field into the QD [12].

A substantial structural modification of the diamond nanotip, formation of an a-C layer and removal of QD structures were initiated after the FE current increase above 1 µA. A common FN emission behavior (Fig. 1b) was revealed for clean a-C surface, as shown in Fig. 2b. In this case, the spectrum consists of a single asymmetric peak which shifts non-linearly with $V$ due to the voltage drop, $IR$, developed across the emitter having an electrical resistance $R$. The FE current follows the standard FN equation $I(V) \sim aV^2 \exp(-b/V)$, where $a$ and $b$ are fitting parameters.

When a carbon nanowire was formed at the top of the a-C layer, the FE was governed by the CB effect (Fig. 1c). As a result, a periodic modulation (Coulomb staircase) in the $I(V)$ curve and sawtooth oscillations in the $\varepsilon_{peak}(V)$ dependence are observed (Fig. 2c). The amplitude of $\varepsilon_{peak}(V)$ oscillations gives the charging energy, which is determined by the total capacitance of the nanowire, $C$, as $\delta\varepsilon = e^2/C$. The period of the oscillations is determined by the nanowire capacitance with respect to the anode electrode, $C_A$, as $\delta V = e/C_A$. The model of FE in the CB regime [17], based on the master equation (see below), reproduces well the experimental data, as shown by the red lines in Fig. 2c.

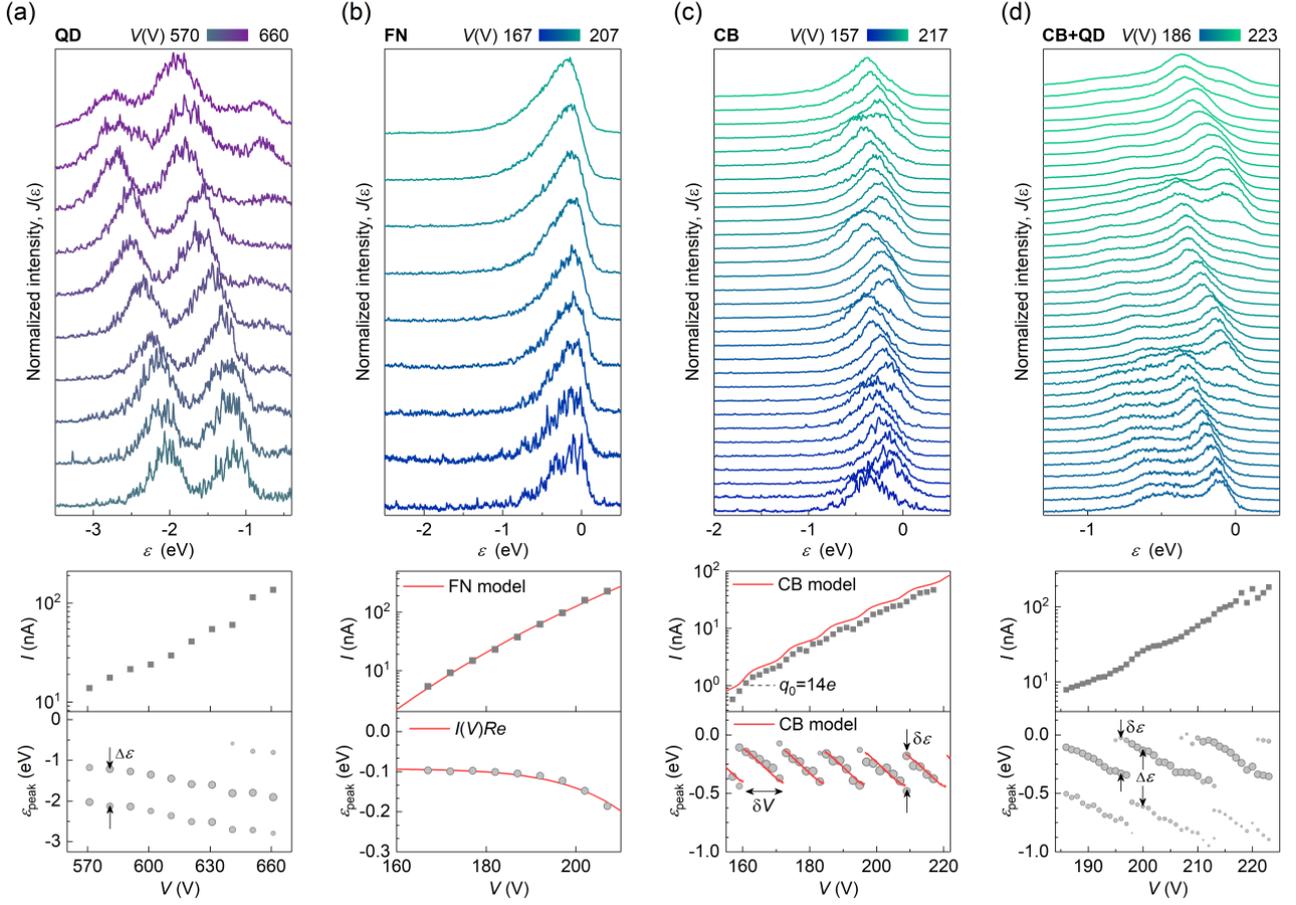

**FIG. 2.** (a-d) $J(\varepsilon)$, $I$ and $\varepsilon_{peak}$ as a functions of $V$, measured at different stages of the FE-induced reconstruction shown in Fig. 1 (a-d), correspondingly. Each $J(\varepsilon)$ curve is normalized to unity. The size of each point in $\varepsilon_{peak}(V)$ is proportional to the amplitude of the energy peak, obtained by $J(\varepsilon)$ curve fitting [17]. The red lines in (b) are the $I(V)$ curve fit by the FN equation and fit of the shift in $\varepsilon_{peak}(V)$ with $R=400$ k$\Omega$. The red lines in (c) are fits of $I(V)$ and $\varepsilon_{peak}(V)$ dependencies using CB model [17]. The $I(V)$ curve fit is shifted for clarity.

Thus, the CB effect observed for nanowires and the quantum confinement effect observed for QD structures manifest themselves differently in FE, and by measuring the energy spectra, one can unambiguously distinguish between them. Moreover, the spectral features, which are specific to each effect, can appear simultaneously, when a QD structure is formed at the apex of a nanowire. We found that after the QD formation [17], the shape of the spectrum changed from a single (Fig. 2c) to multiple peaks (Fig. 2d), and simultaneous sawtooth oscillations were revealed for two well-separated energy peaks, spaced by $\Delta\varepsilon$. It should be noted, that the $I(V)$ curve in Fig. 2d has a much steeper slope than in Fig. 2c, and the current $I$ is more than twice higher in the high-$V$ region. At the same time, the period $\delta V$ and amplitude $\delta\varepsilon$ of the oscillations are almost the same. This can be well explained within the model shown in Fig. 1d, where electrons tunnel into the QD from the nanowire Fermi level, which oscillates with $V$ due to CB. Since the QD is much smaller in size as compared to the nanowire, the capacitances $C$ and $C_A$, which determine $\delta\varepsilon$ and $\delta V$, did not vary significantly after the QD formation. However, the profile of the triangular barrier and the FE mechanism can be very different in the presence of QD, which explains the observed changes in the $I(V)$ curve.

Next, we consider the case of FE from the small-size nanowires (Fig. 1e), where quantum confinement is significant and can be observed experimentally. The size of a nanowire can be estimated from experimental data by extrapolating the CB oscillations to $V=0$ V, which gives its characteristic charge, $q_0$, as a ratio $q_0 \approx V_0/\delta V$, where $V_0$ is the voltage corresponding to the FE current $I_0$. For example, in Fig. 2c $q_0 \sim 14e$ (at $I_0$=1nA). We found, that the evidence for quantum confinement effects appear at twice smaller $q_0$ values, as demonstrated in Fig. 3a, b for a sample with $q_0 \sim 7e$. It should be noted that in this case, limitations in the spectroscopy resolution did not allow us to resolve discrete energy peaks, as in the case of QDs, because of the smaller value of $\Delta\varepsilon$. Nevertheless, the current-voltage measurements revealed a well-reproducible short-period modulation of the current amplitude in the Coulomb staircase (Fig. 3a). This is clearly visible in the normalized differential conductance dependence, $(dI/dV) / (I/V)$, presented in Fig. 3b, where high-amplitude peaks spaced by $\delta V \sim 25$ V are superimposed by less intense peaks with about threefold smaller spacing. This behavior resembles the modified Coulomb staircases observed in the transport through solid-state semiconducting QDs with strong asymmetry in the barriers, where discrete states and single-electron charging coexist [18,19]. However, in our case, the underlying physics is quite different due to the voltage-dependent resistance of the emission barrier and completely different geometry of the experiment.

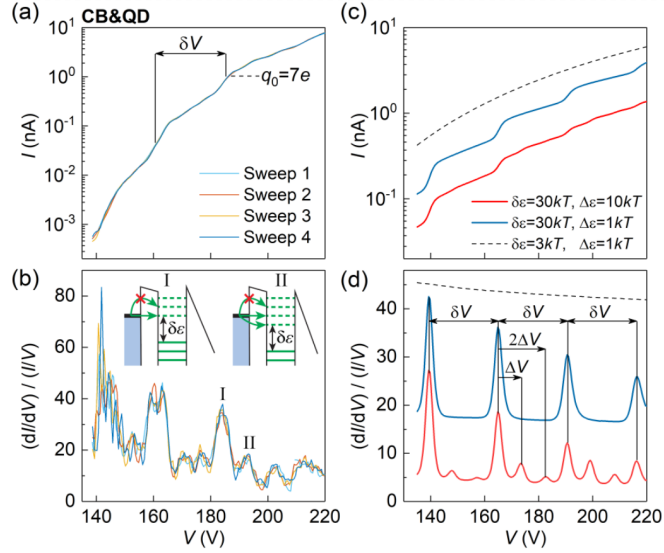

**FIG. 3.** (a) Experimental *I(V)* characteristics and (b) the corresponding normalized differential conductance curves for a small nanowire. Insets in (b) show schematic diagrams of electron tunneling. (c, d) Results of the simulations performed at *T*=300 K for different δε and Δε. See text for details.

In order to model the charge transport in a double-barrier structure with a quantum confined nanowire (Fig. 1e), it is straightforward to use the master equation [3,20], which allows the determination of the probabilities $P_N$ of finding the nanowire in the state with *N* electrons at a fixed voltage. In the stationary case, it can be written as

$$P_N \sum_k \left[ \Gamma_k (1 - f(\varepsilon_k + \Delta U_N)) g_n(\varepsilon_k) + \frac{I_{FE}(\varepsilon_k, F_N)}{e} \right] = P_{N-1} \sum_k \Gamma_k (1 - g_N(\varepsilon_k)) f(\varepsilon_k + \Delta U_N). \quad (1)$$

Here the sums on the left and right sides describe the total tunneling rates from and to the nanowire discrete energy levels, $\varepsilon_k$, respectively. Each tunneling event is associated with a change in the Coulomb energy $\Delta U_N = U_N - U_{N-1} = \delta\varepsilon(N - 1/2 - V/\delta V)$. The function $f(\varepsilon) = 1/(1 + \exp(\varepsilon/k_B T))$ is the Fermi distribution, $g_N(\varepsilon_k)$ is the electron distribution function in the nanowire and $I_{FE}(\varepsilon_k, F_N)$ is the partial FE current from energy level $\varepsilon_k$ at an electric field $F_N = \alpha_V V + \alpha_N N$, where $\alpha_V$ and $\alpha_N$ are parameters determined by the geometry of the nanowire and surrounding electrodes. The solution of Eq. (1) gives the probabilities distribution $P_N$. The total FE current can then be calculated as a sum over all partial currents $I = \sum_N \sum_k I_{FE}(\varepsilon_k, F_N) P_N$.

Good qualitative agreement between the simulation and experiment was obtained even in the simplest case of equidistant energy levels $\varepsilon_k = k\Delta\varepsilon$ populated according to the Fermi distribution $g_n(\varepsilon_k) = f(\varepsilon_k)$, constant tunneling rates $\Gamma_k = \Gamma_0$ = const, and partial FE currents described by the standard Young formula for electron emission from a free-electron gas [15,17].

When the thermal energy, $kT$, is substantially smaller than both energies $\delta\varepsilon$, $\Delta\varepsilon$, the simulated $I(V)$ curve (red line in Fig. 3c) has the staircase shape with an additional modulation that best matches the experiment when $\delta\varepsilon/\Delta\varepsilon \sim 3$. The differential conductance curve exhibits high-amplitude periodic peaks, spaced by $\delta V$, and additional peaks shifted by an integer number of $\Delta V = (\Delta\varepsilon/\delta\varepsilon)\delta V$ from each main peak (Fig. 3d). In the low-$V$ region, the additional peaks are small, and the differential conductance curve is close to the blue curve simulated at $\Delta\varepsilon \sim kT$, i.e. in the classical CB regime. In the high-$V$ region, amplitudes of the main and additional peaks become comparable.

The physical mechanism underlying the conductance behavior is explained schematically in the inset in Fig. 3b. Scheme I corresponds to the main peak in the differential conductance, when just one energy level, that is aligned with the Fermi level in the electron reservoir, is available for tunneling into the nanowire. The levels shift downward with increasing $V$, due to the change in the Coulomb energy $\Delta U_N$. When the second level aligns with the Fermi level of the reservoir (Scheme II), the tunneling probability through the inner barrier (between the reservoir and the nanowire) sharply increases and an additional peak in the differential conductance is observed. With a further increase in $V$, the third level becomes allowed for tunneling, and so on, until the total energy shift reaches $\delta\varepsilon$ and another main peak appears. At low $V$, the additional peaks are weak, since the electron transport is determined by a less transparent outer (triangular) barrier, which depends only on the number of electrons on the nanowire and does not depend on the number of levels available for tunneling. The additional peaks increase with increasing $V$, because the inner barrier becomes a bottleneck for transport. At high $V$, the oscillations are washed out and the differential conductance curve follows the dashed line in Fig. 4d, which is simulated with an order of magnitude smaller $\delta\varepsilon$ and $\Delta\varepsilon$.

It is important to note that in the experiment the differential conductance peaks are not equidistant (Fig. 4b), i.e. $\delta V$ and $\Delta V$ increase with $V$. This can be attributed to the change in capacitive characteristics of the nanowire (in particular $C_A$, since it defines $\delta V$) due to the semiconducting properties of the carbon emitter. Because of the limited carrier concentration, the electric field penetrates into the nanowire and the supporting a-C layer, and the capacitances decrease with increasing voltage. We obtained good agreement with the experimental peak positions by introducing a slight linear decrease in $C_A$ of 12% into the model (see Supplemental Material Fig. S4 [17]).

Let us finally discuss the possible mechanisms of electron tunneling in the considered structures, namely coherent and sequential tunneling processes, which, in general, both contribute to the transport through double-barrier structures [21]. The dominant mechanism depends on the coupling between the reservoir and the emitting nanostructure [22,23]. The coupling strength can be characterized by the tunneling resistance, $R_T$. In the case of FE from nanowires (Fig. 2c), the coupling is weak and $R_T \gg R_q$ [6], where $R_q = h/e^2 \sim 25.8$ k$\Omega$ is the resistance quantum (von Klitzing constant). Under these conditions, single-electron charging effects become important, and transport occurs via sequential tunneling events described by the master equation. In the case of QDs, the coupling is strong and coherent tunneling prevails, which is reflected by the multi-peak structure of the spectra (Fig. 2a). In the intermediate case of small nanowires, the sequential tunneling approximation is still valid and the CB theory is consistent with experiment (Fig. 3). However, the discrepancies in the shape and amplitudes of the differential conductance peaks may indicate an additional contribution of coherent tunneling [23,24].

In conclusion, we studied the combined effect of quantum confinement and single-electron charging in FE from heterostructure nanotips composed of self-assembled carbon QDs and nanowires. The distinctive features, as compared to the typical properties of solid-state QDs, observed in the FE experiments, are explained by the strong dependence of the emission barrier transparency on the applied voltage and the charge of the self-assembled nanostructure, as well as by the peculiar geometry of vacuum FE devices, in which the source and drain are separated by a macroscopic distance. The observation of size and charge quantization effects at room temperature is possible in our experiments owing to the strong localization of a well-defined number of electrons in the carbon nanostructure and, at the same time, strong concentration of the electric field on its surface. This offers further opportunities for using the tip-shaped carbon heterostructures, e.g. in scanning QD microscopy [25] for probing local electrostatic potential fields and in laser-trigged single-electron sources [26,27] for low-energy electron holography [28] and ultrafast electron microscopy [29].

## Acknowledgements

Experimental and theoretical research was supported by Russian Science Foundation (Grant No. 19-72-10067). V.P. and D.L.-H. are grateful for German Federal Ministry of Education and Research under (Project No. 05K13PX2) for infrastructure support. A.N.O. and V.I.K. are also grateful to the Academy of Finland (Grant No. 334552) for travel support.

## Author Contributions

V.I.K. performed the experiments and theoretical calculations, V.P. assisted with electron spectroscopy measurements, V.I.K., D.L.-H. and A.N.O. co-wrote the manuscript.

# Supplemental Material for
# "Coulomb blockade and quantum confinement in field electron emission from heterostructured nanotips"

Victor I. Kleshch[1,*], Vitali Porshyn[2], Dirk Lützenkirchen-Hecht[2], Alexander N. Obraztsov[1,3]

[1]Department of Physics, Lomonosov Moscow State University, Moscow 119991, Russia

[2]Physics Department, Faculty of Mathematics and Natural Sciences, University of Wuppertal, Wuppertal 42119, Germany

[3]Department of Physics and Mathematics, University of Eastern Finland, Joensuu 80101, Finland

* Corresponding author, e-mail: klesch@polly.phys.msu.ru

## I. Initial stabilization of field emission

Field emission (FE) from a pristine (unmodified) diamond needle at low currents about 1nA was very unstable, as it is typically observed for any uncleaned tip during the first FE tests [1-5]. In this initial measurements, the electron emission occurs from nanoscale structures formed at the apex due to the field-induced diffusion of surface atoms. By increasing the current to about 100 nA the nanostructures and electron emission become relatively stable. The same behavior is usually observed e.g. for carbon nanotubes [2]. An example of FE stabilization from a diamond needle is shown in Fig. S1.

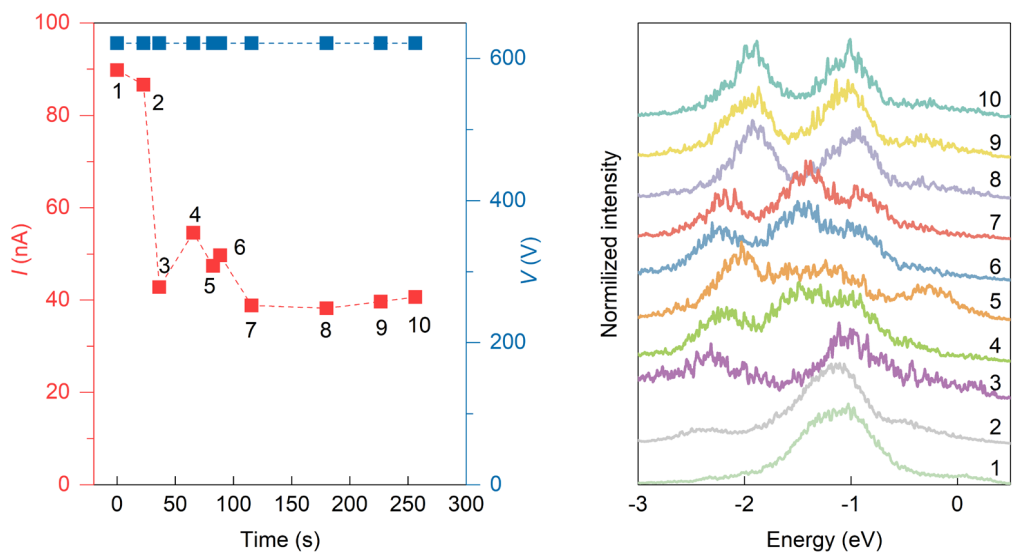

**FIG. S1.** Dependence of the FE current, $I$, on time and the corresponding electron spectra at a fixed voltage, demonstrating the stabilization of the FE characteristics of pristine diamond needle. The measurements shown in Fig. 2a of the main text were performed right after this stabilization.

## II. Fitting the energy spectra

In order to obtain the positions and amplitudes of the peaks in the energy spectra, we use the standard Young formula of total energy distribution for the emission from a free-electron gas [6]:

$$j(\varepsilon, F) = \frac{j_0(F)}{d(F)} \frac{e^{\varepsilon/d(F)}}{1 + e^{\varepsilon/kT}}, \quad (S1)$$

where $\varepsilon$ is the kinetic electron energy relative to the Fermi level, $F$ is the electric field, $j_0$ is the total current density at $T=0$ K, and $d$ is the function of $F$ determined in Section IV.

Each peak in the spectrum is fitted using Eq. (S1) at $F=$const, i.e. by the function

$$j_{\text{fit}}(\varepsilon) = \frac{C_1 e^{C_2(\varepsilon - \varepsilon_{peak})}}{1 + e^{C_3(\varepsilon - \varepsilon_{peak})}}, \quad (S2)$$

where $\varepsilon_{peak}$, $C_1$, $C_2$, $C_3$ are the fitting coefficients. The maximum of $j_{\text{fit}}(\varepsilon)$ gives the amplitude of the peak and $\varepsilon_{peak}$ gives its position. Figure S2 shows examples of the spectra fits for each plot in Fig. 2 of the main text.

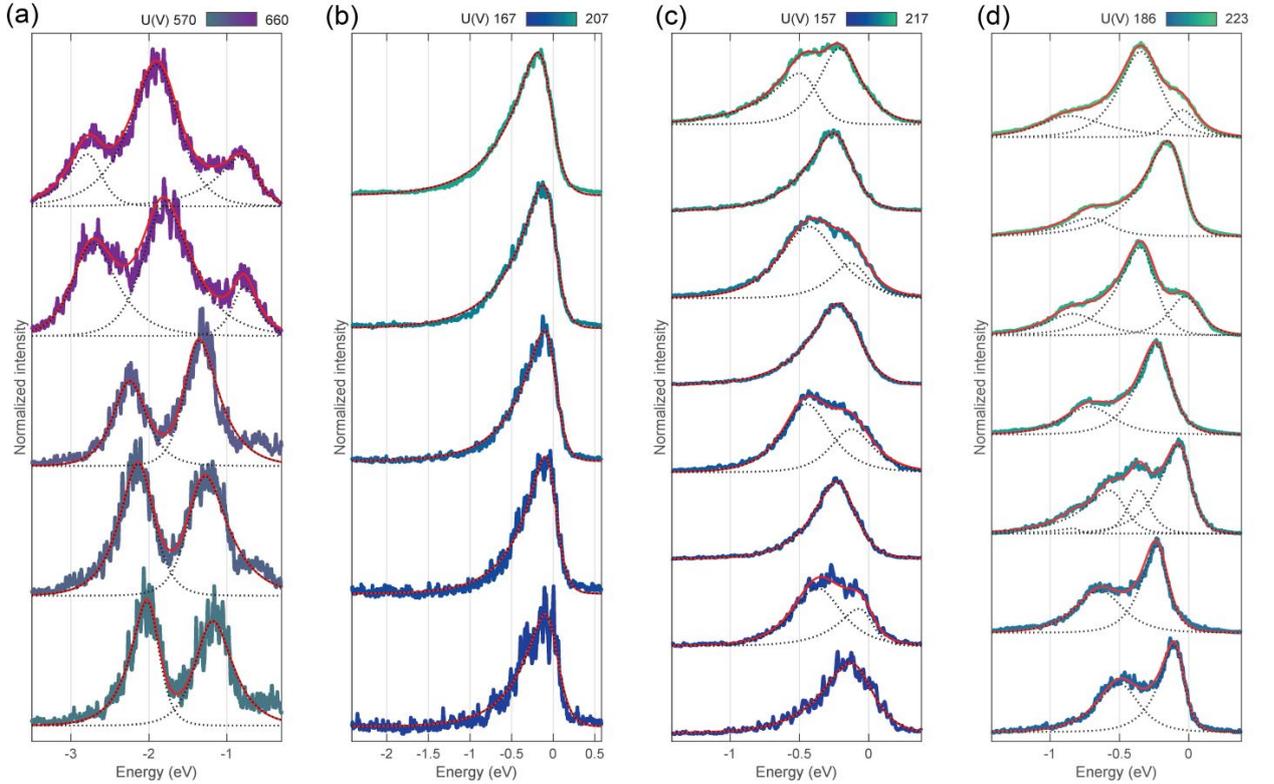

**FIG. S2.** Fits of the energy spectra presented in Fig. 2 of the main text. The dashed lines show fits of individual peaks by Eq. (S2). The resulting fits are shown by the red solid lines.

## III. QD-nanowire structure formation

Similar to the case of pristine diamond needles, QD structures can sometimes be formed during FE from nanowires, since adsorption layers accumulate over time even in ultra-high-vacuum conditions [2]. An example of the spectrum structure transformation from a single-peak to multiple-peak as a result of the QD formation is shown in Fig. S3. It is worth noting, that the FE characteristics of QD structures become unstable at currents above 100 nA, as can be seen in the $I(V)$ curve in Fig. 2d of the main text.

With a further increase in the current, QDs are completely removed, while the nanowires can remain stable up to currents of several microamperes [7].

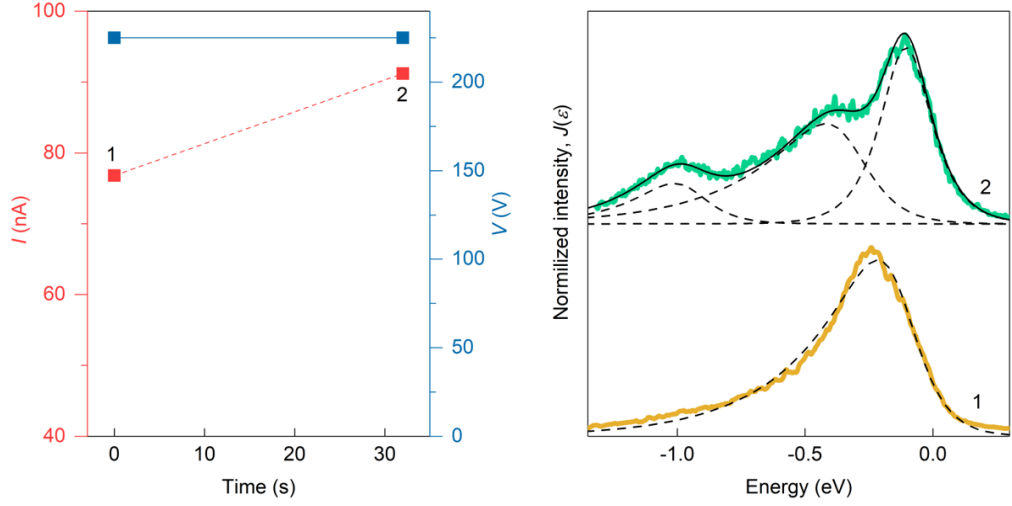

**FIG. S3.** Two successive measurements of electron spectrum and FE current at a fixed voltage, demonstrating the formation of a QD at the top of the nanowire. These measurements were performed in between the measurements shown in Fig. 2c and Fig. 2d of the main text. The dashed black lines are the fits of individual peaks in the spectra. The solid black line is the fit of the spectrum 2, which consists of three distinct peaks.

### IV. Transport model

Charge transport in a double-barrier structure is determined by the master equation given by Eq. (1) in the main text. The transfer of electrons through the barrier between the nanowire and the support tip is described as tunneling between the electron reservoir and discrete energy levels $\varepsilon_k$ [8] and is associated with the tunneling rate $\Gamma_0$. The transport through the emission barrier is described by the partial FE currents, $I_{FE}$, for tunneling from levels $\varepsilon_k$, given by $I_{FE}(\varepsilon_k, F_N) = K_j \Delta\varepsilon\, j(\varepsilon_k, F_N)$, where $\Delta\varepsilon$ is energy level separation; $K_j$ is the pre-exponential factor, which is associated with the emission area, geometry and the electronic properties of the emitter; $j(\varepsilon_k, F_N)$ is given by Eq. (S1) with the electric field $F_N$ defined in the main text and associated with fitting parameters $\alpha_V$ and $\alpha_N$. The emission current density $j_0(F)$ and the function $d(F)$ in Eq. (S1) were calculated using the standard equations for the emission from a free-electron gas [6]:

$$j_0(F) = AF^2 \exp(-B/F)\ (A/m^2), \qquad (S3)$$

and

$$d(F) = 9.76 \times 10^{-11} F \varphi^{-1/2} t^{-1}(y)\ (eV).$$

Here $A = 1.54 \times 10^{-6} \varphi^{-1} t^{-2}(y)$, $B = 6.831 \times 10^{-9} \varphi^{3/2} v(y)$, $v(y)$ and $t(y)$ are tabulated functions of the variable $y = 3.7495 \times 10^{-5} F^{1/2}/\varphi$, $F$ is in V/m, $\varphi$ is in eV. The work function was set to $\varphi = 5$ eV, which is a typical value for graphite.

The simulation results presented in Fig. 3c, d were obtained with the following parameters: $C_A = 6.3 \times 10^{-21}$ F, $\Gamma_0 = 4.8 \times 10^9$ s$^{-1}$, $k_J = 1.3 \times 10^{-19}$ cm$^2$, $\alpha_V = 4.5 \times 10^7$ m$^{-1}$, $\alpha_N = 8.2 \times 10^8$ V/m. The parameters $\delta\varepsilon$, $\Delta\varepsilon$, and $T$ are defined in the main text. Figure S4 shows a simulation of the differential conductance in the case when linear decrease in $C_A$ with voltage is introduced to the model.

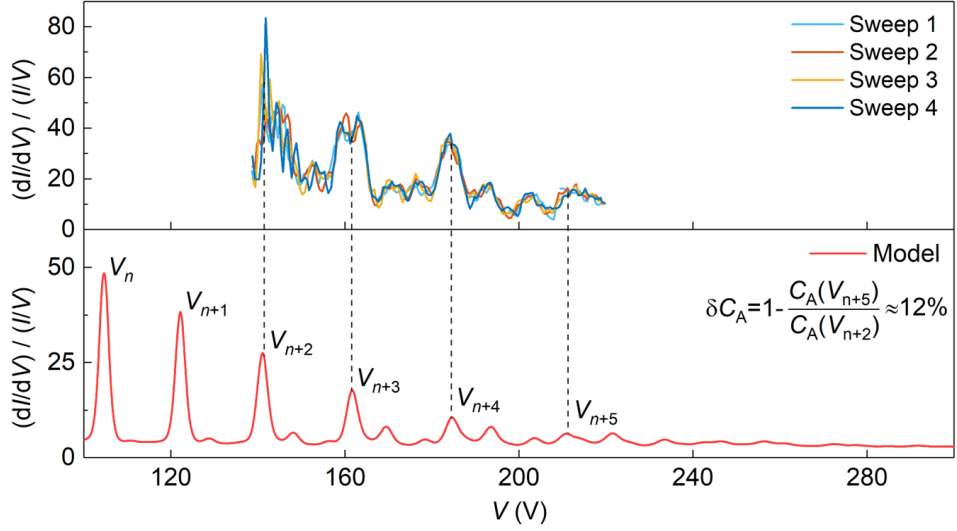

**FIG. S4.** Experimental (top panel) and simulated (bottom panel) dependencies of the normalized differential conductance on the applied voltage, $V$, for a small nanowire. The simulation curve was obtained using the same parameters as in Fig. 3d (red curve) of the main text, except for $C_A$ (the nanowire capacitance with respect to the anode), which is linearly dependent on the voltage as $C_A(V) = C_{A0} - k_C V$, where $C_{A0} = 1.33 \times 10^{-20}$ F and $k_C = 1.82 \times 10^{-23}$ F/V. The relative change of the $C_A$ between voltages $V_{n+2}$ and $V_{n+5}$ is $\delta C_A \approx 12\%$.

It should be noted that Eq. S3 is valid for the FE from a flat surface and can lead to errors when the radius of the emitter becomes the same order as the width of the energy barrier. This explains the discrepancy in the absolute values of current and differential conductance between the experiment and simulation in Fig. 3 of the main text. We used Eq. S3 to simulate the behavior of the current in the most general case, since the exact atomic and electronic structure of the confined nanowire is not known, and, therefore, the model can give only qualitative agreement with experiment. In the case of weak confinement, quantitative agreement can be achieved, as it was shown in our previous work [7], where simulations were performed using a simpler model, which is a special case of the present model with $\delta\varepsilon \gg \Delta\varepsilon \sim kT$. The fits in Fig. 2c of the main text were obtained using this simplified model with the following parameters: $C_A = 13.5 \times 10^{-21}$ F, $C = 5.1 \times 10^{-19}$ F, $R = 200$ kΩ (tunneling resistance), $\alpha_V = 4.5 \times 10^7$ m$^{-1}$, $\alpha_N = 8.2 \times 10^8$ V/m, $T = 300$ K.